\documentstyle[12pt]{article}
\newcommand{\n}{\hspace*{-2.5mm}}
\newcommand{\gsim}{\;\rlap{\lower 3.5 pt \hbox{$\mathchar \sim$}} \raise 1pt
 \hbox {$>$}\;}
\newcommand{\lsim}{\;\rlap{\lower 3.5 pt \hbox{$\mathchar \sim$}} \raise 1pt
 \hbox {$<$}\;}

\begin{document}
\title{\vskip-3cm{\baselineskip14pt
\centerline{\normalsize DESY 95--008\hfill}
\centerline{\normalsize MPI/PhT/95--3\hfill}
\centerline{\normalsize hep-ph/9501392 \hfill}
\centerline{\normalsize January 1995\hfill}
}
\vskip1.5cm
Two-Loop ${\cal O}(\alpha_sG_Fm_t^2)$ Corrections to Higgs Production at LEP}
\author{Bernd A. Kniehl\\
Max-Planck-Institut f\"ur Physik, Werner-Heisenberg-Institut\\
F\"ohringer Ring 6, 80805 Munich, Germany\\ \\
Michael Spira\\
II. Institut f\"ur Theoretische Physik\thanks{Supported
by Bundesministerium f\"ur Forschung und Technologie, Bonn, Germany,
under Contract 05~6~HH~93P~(5),
and by EEC Program {\it Human Capital and Mobility} through Network
{\it Physics at High Energy Colliders} under Contract
CHRX-CT93-0357 (DG12 COMA).}, Universit\"at Hamburg\\
Luruper Chaussee 149, 22761 Hamburg, Germany}
\date{}
\maketitle
\begin{abstract}
We evaluate the two-loop ${\cal O}(\alpha_sG_Fm_t^2)$ correction to the $ZZH$
coupling in the Standard Model by means of a low-energy theorem, assuming that
the top quark is much heavier than the Higgs boson.
We then construct a heavy-top-quark effective Lagrangian for the $ZZH$
interaction that accommodates the ${\cal O}(G_Fm_t^2)$ and
${\cal O}(\alpha_sG_Fm_t^2)$ corrections and derive from it the corresponding
corrections to the $H\to ZZ$ decay as well as those to Higgs-boson production
at LEP1, via $Z\to f\bar fH$, and at LEP2, via $e^+e^-\to ZH$.
In all cases, the leading ${\cal O}(G_Fm_t^2)$ terms are considerably
screened by their QCD corrections, if the on-shell renormalization scheme
with $G_F$ as a basic parameter is employed.
\end{abstract}

\section{Introduction}

The Higgs boson is the missing link in the Standard Model (SM).
The discovery of this particle and the study of its characteristics are among
the prime goals of present and future high-energy colliding-beam experiments.
At LEP1 and SLC, the Higgs boson is currently being searched for in the
decay products of the $Z\to f\bar fH$ channel \cite{bjo}.
At the present time, the failure of this search allows one to rule out the
mass range $M_H\le63.9$~GeV at the 95\% confidence level \cite{jan}.
At LEP2, the quest for the Higgs boson will be continued using the
Higgs-strahlung mechanism, $e^+e^-\to ZH$ \cite{ell,iof}.

Once a novel scalar particle is discovered, it will be crucial to decide if it
is the very Higgs boson of the SM or if it lives in some more extended Higgs
sector.
To that end, precise knowledge of the SM prediction will be mandatory,
i.e., quantum corrections must be taken into account.
The current knowledge of radiative corrections to the production and decay
processes of the SM Higgs boson has recently been reviewed \cite{bak}.
Heavy-fermion effects on $\Gamma\left(Z\to f\bar fH\right)$ and
$\sigma(e^+e^-\to ZH)$, at one loop, have been analyzed in Ref.~\cite{fle}.
The full one-loop electroweak correction to $\Gamma\left(Z\to f\bar fH\right)$
has been presented in Ref.~\cite{zffh}.
The one-loop QED \cite{kle,ffzh} and weak \cite{ffzh,fj} corrections to
$\sigma(e^+e^-\to ZH)$ are also available.
The theoretical predictions for $\sigma(e^+e^-\to ZH)$ at LEP2 energies have
recently been collected and updated \cite{gkw}.

In view of experimental evidence for a heavy top quark, with
$m_t=(174\pm16)$~GeV \cite{cdf}, the $m_t$-dependent corrections are
particularly important.
In the case of $\Gamma\left(Z\to f\bar fH\right)$ and $\sigma(e^+e^-\to ZH)$,
they may be accommodated by multiplying the respective Born formulae with
$(1+\Delta_{f\bar fZH})$, where \cite{fle,zffh,ffzh}
\begin{equation}
\label{one}
\Delta_{f\bar fZH}
=N_cx_t\left(-{2\over3}-{8c_w^2Q_fv_f\over v_f^2+a_f^2}\right),
\end{equation}
which is negative for all flavours.
Here, $N_c=3$, $x_t=\left(G_Fm_t^2/8\pi^2\sqrt2\,\right)$,
$c_w^2=1-s_w^2=M_W^2/M_Z^2$,
$v_f=2I_f-4s_w^2Q_f$, $a_f=2I_f$,
$Q_f$ is the electric charge of $f$ in units of the positron charge,
$I_f$ is the third component of weak isospin of the left-handed component
of $f$,
and it is understood that the Born results are expressed in terms of the Fermi
constant, $G_F$.
In the case of $\sigma(e^+e^-\to ZH)$, $f=e$ in Eq.~(\ref{one}).

The goal of this paper is to derive the QCD correction to Eq.~(\ref{one}).
This will be achieved by means of an appropriate low-energy theorem
\cite{ell,vai,daw}.
Generally speaking, this theorem relates the amplitudes of two processes which
differ by the insertion of an external Higgs-boson line carrying zero momentum.
It provides a convenient tool for estimating the properties of a Higgs boson
which is light compared to the loop particles.
In the literature, a similar theorem has been applied to derive low-$M_H$
effective Lagrangians for the $\gamma\gamma H$ and $ggH$ interactions at one
\cite{vai} and two loops \cite{daw}.
In a previous paper \cite{hbb}, we have employed this theorem to find
the non-universal ${\cal O}(\alpha_sG_Fm_t^2)$ correction to
$\Gamma\left(H\to b\bar b\,\right)$.

For the reader's convenience, we shall review the gist of the matter.
This low-energy theorem may be derived by observing that the
interactions of the Higgs boson with the massive particles
in the SM emerge from their mass terms by substituting
$m_i\to m_i(1+H/v)$, where $m_i$ is the mass of the respective particle,
$H$ is the Higgs field, and $v$ is the Higgs vacuum expectation value.
On the other hand, a Higgs boson with zero momentum is represented by a
constant
field, since $i\partial_\mu H=[P_\mu,H]=0$, where $P_\mu$ is the
four-momentum operator.
This immediately implies that a zero-momentum Higgs boson may be attached
to an amplitude, ${\cal M}(A\to B)$, by carrying out the operation
\begin{equation}
\label{let}
\lim_{p_H\to0}{\cal M}(A\to B+H)={1\over v}\sum_i
{m_i\partial\over\partial m_i}{\cal M}(A\to B),
\end{equation}
where $i$ runs over all massive particles which are involved in the transition
$A\to B$.
Here, it is understood that the differential operator does not act
on the $m_i$ appearing in coupling constants, since this would generate
tree-level interactions involving the Higgs boson that do not exist in the SM.
Special care must be exercised if this low-energy theorem is to be applied
beyond the leading order.
Then it must be formulated for the bare quantities of the theory.
The renormalization is performed after the left-hand side of Eq.~(\ref{let})
has been evaluated.

This paper is organized as follows.
In Sect.~2, we derive a heavy-top-quark effective Lagrangian for the $ZZH$
interaction, including the ${\cal O}(G_Fm_t^2)$ and
${\cal O}(\alpha_sG_Fm_t^2)$ corrections, by means of the low-energy theorem.
{}From this Lagrangian we may instantly read off the ${\cal
O}(\alpha_sG_Fm_t^2)$
correction to $\Gamma(H\to ZZ)$, which, to our knowledge, is a new result.
In order to obtain the corresponding corrections to
$\Gamma\left(Z\to f\bar fH\right)$ and $\sigma(e^+e^-\to ZH)$,
we also need to include similar corrections originating in the gauge sector.
This will be done in Sect.~3 by invoking the so-called improved Born
approximation (IBA) \cite{iba}.
Section~4 contains our conclusions.

\section{Effective Lagrangian}

We use dimensional regularization in $n=4-2\epsilon$ space-time dimensions and
introduce a 't~Hooft mass, $\mu$, to keep the coupling constants dimensionless.
As usual, we take $\gamma_5$ to be anticommuting.
We work in the on-shell renormalization scheme \cite{aok}, with $G_F$ as a
basic parameter.

Prior to actually evaluating loop amplitudes, we develop the general formalism.
The starting point of our analysis is the amplitude characterizing the
propagation of an on-shell $Z$ boson in the presence of quantum effects due
to a virtual high-mass top quark,
\begin{equation}
\label{zz}
{\cal M}(Z\to Z)=(M_Z^0)^2-\left.\Pi_{ZZ}(q^2)\right|_{q^2=(M_Z^0)^2},
\end{equation}
where $\Pi_{ZZ}(q^2)$ is the unrenormalized transverse self-energy of the $Z$
boson, with momentum $q$, and is expressed in terms of bare parameters.
Here and in the following, bare parameters are marked by the superscript 0.
In the $G_F$ representation, $\Pi_{ZZ}(q^2)$ is proportional to $(M_Z^0)^2$,
which originates from the two $t\bar tZ$ gauge couplings.
Apart from this prefactor, we may put $q^2=0$ in Eq.~(\ref{zz}),
since we are working in the high-$m_t$ approximation.
The low-energy theorem (\ref{let}) now tells us that we may attach a
zero-momentum Higgs boson to the $Z\to Z$ transition amplitude by carrying
out the operation
\begin{equation}
\lim_{p_H\to0}{\cal M}(Z\to Z+H)={1\over v^0}\left(
{m_t^0\partial\over\partial m_t^0}+{M_Z^0\partial\over\partial M_Z^0}\right)
{\cal M}(Z\to Z),
\end{equation}
where we must treat the overall factor $(M_Z^0)^2$ of $\Pi_{ZZ}(0)$
in Eq.~(\ref{zz}) as a constant.
This leads us to
\begin{equation}
\lim_{p_H\to0}{\cal M}(Z\to Z+H)={2(M_Z^0)^2\over v^0}(1+E),
\end{equation}
with
\begin{equation}
\label{ezzh}
E=-{(m_t^0)^2\partial\over\partial(m_t^0)^2}\,
{\Pi_{ZZ}(0)\over (M_Z^0)^2}.
\end{equation}

We are now in the position to write down the heavy-top-quark effective
$ZZH$ interaction Lagrangian,
\begin{equation}
\label{bare}
{\cal L}_{ZZH}=(M_Z^0)^2(Z^0)_\mu(Z^0)^\mu{H^0\over v^0}(1+E).
\end{equation}
Then, we have to carry out the renormalization procedure, i.e.,
we have to split the bare parameters into renormalized ones and counterterms.
We fix the counterterms according to the on-shell scheme.
In the case of the $Z$-boson mass and wave function, we have
\begin{eqnarray}
(M_Z^0)^2&\n=\n&M_Z^2+\delta M_Z^2,\nonumber\\
(Z^0)^\mu&\n=\n&(1+\delta Z_Z)^{1/2}Z^\mu,
\end{eqnarray}
with
\begin{eqnarray}
\delta M_Z^2&\n=\n&\Pi_{ZZ}(0),\nonumber\\
\delta Z_Z&\n=\n&-\Pi_{ZZ}^\prime(0),
\end{eqnarray}
where we have neglected $M_Z$ against $m_t$ in the loop amplitudes.
For dimensional reasons, $\delta Z_Z$ does not receive corrections in
${\cal O}(G_Fm_t^2)$ and ${\cal O}(\alpha_sG_Fm_t^2)$.
{}From the analysis of $\Gamma\left(H\to f\bar f\right)$ in ${\cal
O}(G_Fm_t^2)$
and ${\cal O}(\alpha_sG_Fm_t^2)$ \cite{hff} we know that
\begin{equation}
{H^0\over v^0}=2^{1/4}G_F^{1/2}H(1+\delta_u),
\end{equation}
with
\begin{equation}
\label{duni}
\delta_u=N_cx_t\left[{7\over6}-C_F{\alpha_s\over\pi}
\left({\zeta(2)\over2}+{3\over4}\right)\right],
\end{equation}
where $C_F=(N_c^2-1)/(2N_c)=4/3$.
$\delta_u$ is a universal correction, which occurs as a building block in the
renormalizations of all Higgs-boson production and decay processes.
Putting everything together, we obtain the renormalized version of
Eq.~(\ref{bare}),
\begin{equation}
\label{reno}
{\cal L}_{ZZH}=2^{1/4}G_F^{1/2}M_Z^2Z_\mu Z^\mu H(1+\delta_{ZZH}),
\end{equation}
with
\begin{equation}
\delta_{ZZH}=\delta_u+{\delta M_Z^2\over M_Z^2}+E.
\end{equation}
In order for $\delta_{ZZH}$ to be finite through ${\cal O}(\alpha_sG_Fm_t^2)$,
we still need to renormalize the top-quark mass in the ${\cal O}(G_Fm_t^2)$
expressions for $\delta M_Z^2/M_Z^2$ and $E$, i.e., we need to
substitute
\begin{equation}
m_t^0=m_t+\delta m_t,
\end{equation}
with \cite{hff,tar}
\begin{equation}
\label{top}
{\delta m_t\over m_t}=-{\alpha_s\over4\pi}C_F
\left({4\pi\mu^2\over m_t^2}\right)^\epsilon
\Gamma(1+\epsilon){3-2\epsilon\over\epsilon(1-2\epsilon)},
\end{equation}
where $\Gamma$ is Euler's gamma function.

Now, we turn to the evaluation of the two-loop amplitudes.
To simplify the notation, we introduce $t=m_t^2$ and $Z=\delta M_Z^2/M_Z^2$.
We label ${\cal O}(G_Fm_t^2)$ and ${\cal O}(\alpha_sG_Fm_t^2)$ contributions
with the subscripts 1 and 2, respectively.
Quantities with (without) the superscript 0 are written in terms of
$m_t^0$ ($m_t$).
First, we shall list the ${\cal O}(G_Fm_t^2)$ results.
These may be extracted from Ref.~\cite{hzz} and read
\begin{eqnarray}
\label{zzenro}
Z_1&\n=\n&N_cx_t\left({4\pi\mu^2\over m_t^2}\right)^\epsilon
\Gamma(1+\epsilon)\left(-{2\over\epsilon}+{\cal O}(\epsilon)\right),\\
\label{ezenro}
E_1&\n=\n&N_cx_t\left({4\pi\mu^2\over m_t^2}\right)^\epsilon
\Gamma(1+\epsilon)\left({2\over\epsilon}-2+{\cal O}(\epsilon)\right).
\end{eqnarray}
In Ref.~\cite{hzz}, $E_1$ has been computed diagrammatically.
This allows us to check Eq.~(\ref{ezzh}) in ${\cal O}(G_Fm_t^2)$.
In fact, one immediately verifies that
\begin{equation}
\label{eone}
E_1=-{t\partial\over\partial t}Z_1.
\end{equation}
The QCD corrections to the electroweak-gauge-boson vacuum polarizations have
been evaluated by means of dispersion relations in Ref.~\cite{vac}.
These results may be converted to dimensional regularization by adjusting the
ultraviolet regulators as described in Refs.~\cite{hff,thr}.
In the case of $Z_2$, this leads to
\begin{equation}
\label{abdel}
Z_2=N_cC_F{\alpha_s\over\pi}x_t
\left({4\pi\mu^2\over m_t^2}\right)^{2\epsilon}\Gamma^2(1+\epsilon)
\left({3\over2\epsilon^2}+{11\over4\epsilon}+{31\over8}+{\cal O}(\epsilon)
\right),
\end{equation}
which agrees with the result obtained in last two papers of Ref.~\cite{djo}.
Notice that Eq.~(\ref{abdel}) already contains the contribution proportional
to $\delta m_t$ which emerges from the renormalization of the top-quark mass in
Eq.~(\ref{zzenro}).
Our final aim is to compute $E_2$.
According to Eq.~(\ref{ezzh}), we have
\begin{equation}
\label{ebaretwo}
E_2^0=-{t\partial\over\partial t}Z_2^0.
\end{equation}
Furthermore, we have
\begin{eqnarray}
\label{ztwo}
Z_2&\n=\n&Z_2^0+\delta Z_2,\\
E_2&\n=\n&E_2^0+\delta E_2,
\end{eqnarray}
where the counterterms are obtained by scaling the one-loop results,
\begin{eqnarray}
\label{dztwo}
\delta Z_2&\n=\n&{\delta t\over t}\,{t\partial\over\partial t}Z_1,\\
\label{detwo}
\delta E_2&\n=\n&{\delta t\over t}\,{t\partial\over\partial t}E_1.
\end{eqnarray}
Substituting Eq.~(\ref{ztwo}) in Eq.~(\ref{ebaretwo}) and using
Eq.~(\ref{dztwo}), we obtain
\begin{equation}
\label{etwozero}
E_2^0=-{t\partial\over\partial t}Z_2+{t\partial\over\partial t}
\left({\delta t\over t}\,{t\partial\over\partial t}Z_1\right).
\end{equation}
On the other hand, inserting Eq.~(\ref{eone}) into Eq.~(\ref{detwo}) yields
\begin{equation}
\label{deltaetwo}
\delta E_2=-{\delta t\over t}\left({t\partial\over\partial t}\right)^2Z_1.
\end{equation}
Now, combining Eqs.~(\ref{etwozero}) and (\ref{deltaetwo}), we obtain
\begin{equation}
E_2=-{t\partial\over\partial t}Z_2
+\left({t\partial\over\partial t}\,{\delta t\over t}\right)
{t\partial\over\partial t}Z_1.
\end{equation}
Using Eq.~(\ref{eone}) together with
\begin{equation}
{t\partial\over\partial t}\,{\delta t\over t}=-\epsilon{\delta t\over t},
\end{equation}
which may be gleaned from Eq.~(\ref{top}),
this becomes
\begin{equation}
\label{eps}
E_2=-{t\partial\over\partial t}Z_2+\epsilon{\delta t\over t}E_1.
\end{equation}
Obviously, knowledge of the ${\cal O}(\epsilon)$ term of $E_1$ is not
necessary for our purposes.
Using Eqs.~(\ref{top}), (\ref{ezenro}), and (\ref{abdel}), we obtain from
Eq.~(\ref{eps}) the desired two-loop three-point amplitude,
\begin{equation}
\label{etwofinal}
E_2=N_cC_F{\alpha_s\over\pi}x_t
\left({4\pi\mu^2\over m_t^2}\right)^{2\epsilon}\Gamma^2(1+\epsilon)
\left(-{3\over2\epsilon^2}-{11\over4\epsilon}+{5\over8}+{\cal O}(\epsilon)
\right).
\end{equation}
Finally, adding up Eqs.~(\ref{duni}), (\ref{zzenro}), (\ref{ezenro}),
(\ref{abdel}), and (\ref{etwofinal}), we find the ultraviolet-finite
correction in Eq.~(\ref{reno}),
\begin{eqnarray}
\label{dzzh}
\delta_{ZZH}&\n=\n&N_cx_t\left[-{5\over6}+C_F{\alpha_s\over\pi}
\left(-{\zeta(2)\over2}+{15\over4}\right)\right]\nonumber\\
&\n\approx\n&-{5\over2}x_t\left(1-4.684\,{\alpha_s\over\pi}\right).
\end{eqnarray}
This completes the derivation of the effective $ZZH$ interaction Lagrangian.
We recover the notion that, in the electroweak on-shell renormalization scheme
implemented with $G_F$, the ${\cal O}(G_Fm_t^2)$ terms generally are screened
by their QCD corrections.
We are not aware of any counterexample to this rule in electroweak physics.
In the present case, the reduction in size amounts to approximately $-16\%$ for
$\alpha_s=0.108$, which is the value of $\alpha_s(\mu)$ at $\mu=m_t=174$~GeV
if $\alpha_s(M_Z)=0.118$ \cite{bet}.

As a corollary, we note that, in the $G_F$ formulation of the on-shell scheme,
the ${\cal O}(G_Fm_t^2)$ and ${\cal O}(\alpha_sG_Fm_t^2)$ corrections to
$\Gamma(H\to ZZ)$ appear as the overall factor $(1+\delta_{ZZH})^2$.
This reproduces the well-known one-loop result \cite{hzz}.
Of course, in order for the high-$m_t$ approximation to be valid in this case,
$2M_Z<M_H\ll m_t$ must be satisfied, which is probably not very realistic.

\section{Higgs production at LEP}

Armed with the high-$m_t$ effective $ZZH$ interaction Lagrangian derived in
the previous section, we now proceed to the analysis of the two-loop
${\cal O}(\alpha_sG_Fm_t^2)$ corrections to
$\Gamma\left(Z\to f\bar fH\right)$ and $\sigma(e^+e^-\to ZH)$.
Detailed inspection of the one-loop results \cite{zffh,ffzh} reveals that,
apart from the $ZZH$ vertex, leading high-$m_t$ corrections also originate in
the renormalizations of the $Z$-boson propagator and the $Z\gamma$ mixing
amplitude.
The loop-induced $Z\gamma H$ coupling does not produce such a correction,
since it does not involve $t\bar tZ$ axial couplings.
Furthermore, the top quark does not yet enter the residual vertex and box
corrections at one loop, so that ${\cal O}(\alpha_sG_Fm_t^2)$ corrections
do not arise here either.

The so-called improved Born approximation (IBA) \cite{iba} provides a
systematic and convenient method to incorporate the leading high-$m_t$
corrections to processes within the gauge sector of the SM.
The recipe is as follows.
Starting from the Born formula expressed in terms of $c_w$, $s_w$, and the
fine-structure constant defined in Thomson scattering, $\alpha$, one
substitutes
\begin{equation}
\alpha\to\overline\alpha={\alpha\over1-\Delta\alpha},\qquad
c_w^2\to\overline c_w^2=1-\overline s_w^2=c_w^2(1-\Delta\rho),
\end{equation}
where $\overline\alpha$ is the effective fine-structure constant measured at
the $Z$-boson scale and $\Delta\rho=N_cx_t$ is the shift in the $\rho$
parameter induced by the top quark.
To eliminate $\overline\alpha$ in favour of $G_F$, one exploits the relation
\begin{equation}
{\overline\alpha\over\overline c_w^2\overline s_w^2}
={\sqrt2\over\pi}G_FM_Z^2{1\over1-\Delta\rho},
\end{equation}
which correctly accounts for the dominant $m_t$ power terms as well as the
leading logarithms that trigger the running of the fine-structure constant.

In Ref.~\cite{zffh}, it has been explained how the IBA may be combined with
specific knowledge of the high-$m_t$ behaviour of the $ZZH$ vertex \cite{hzz}
to find the ${\cal O}(G_Fm_t^2)$ correction to
$\Gamma\left(Z\to f\bar fH\right)$.
Specifically, the correction factor relative to the Born formula of
$\Gamma\left(Z\to f\bar fH\right)$ written with $G_F$, as in Eqs.~(2) and (3)
of Ref.~\cite{zffh},\footnote{We use this opportunity to correct a misprint in
the published version of Ref.~\cite{zffh}, which is absent in the preprint.
The factor $\sqrt{x/4-x}$ in the last term of Eq.~(3) should be replaced by
$\sqrt{x/(4-x)}$.} is given by
\begin{eqnarray}
\label{dffzh}
1+\Delta_{f\bar fZH}&\n=\n&(1+\delta_{ZZH})^2
{\overline\alpha/\left(\overline c_w^2\overline s_w^2\right)\over
\sqrt2G_FM_Z^2/\pi}\,{\overline v_f^2+a_f^2\over v_f^2+a_f^2}\nonumber\\
&\n=\n&1+2\delta_{ZZH}+
\left(1-{8c_w^2Q_fv_f\over v_f^2+a_f^2}\right)\Delta\rho,
\end{eqnarray}
where $\overline v_f=2I_f-4\overline s_w^2Q_f$, and we have omitted terms of
${\cal O}(G_F^2m_t^4)$ in the second line.
In fact, this reproduces Eq.~(\ref{one}).
Equation~(\ref{dffzh}), with $f=e$, is also the correct ${\cal O}(G_Fm_t^2)$
correction factor for $\sigma(e^+e^-\to ZH)$ \cite{fle,ffzh},
provided that the Born result is written with $G_F$, as in Eq.~(4.7) of
Ref.~\cite{ffzh}.

This procedure readily carries over to ${\cal O}(\alpha_sG_Fm_t^2)$.
We just need to include in Eq.~(\ref{dffzh}) the corresponding terms of
$\delta_{ZZH}$ and $\Delta\rho$.
In the case of $\Delta\rho$, one has \cite{vac,djo}
\begin{equation}
\label{drho}
\Delta\rho
=N_cx_t\left[1-C_F{\alpha_s\over\pi}\left(\zeta(2)+{1\over2}\right)\right].
\end{equation}
Inserting Eqs.~(\ref{dzzh}) and (\ref{drho}) into Eq.~(\ref{dffzh}), we obtain
our final result,
\begin{eqnarray}
\Delta_{f\bar fZH}&\n=\n&N_cx_t\left\{
-{2\over3}\left[1+C_F{\alpha_s\over\pi}\left(3\zeta(2)-{21\over2}\right)\right]
-{8c_w^2Q_fv_f\over v_f^2+a_f^2}
\left[1-C_F{\alpha_s\over\pi}\left(\zeta(2)+{1\over2}\right)\right]\right\}
\nonumber\\
&\n\approx\n&-2x_t\left[1-7.420\,{\alpha_s\over\pi}
+{12c_w^2|Q_f|(1-4s_w^2|Q_f|)\over(1-4s_w^2|Q_f|)^2+1}
\left(1-2.860\,{\alpha_s\over\pi}\right)\right].
\end{eqnarray}
Using $M_W=80.24$~GeV, $M_Z=91.19$~GeV, and $\alpha_s=0.108$, we find that
strong-interaction effects modify the magnitude of $\Delta_{f\bar fZH}$ in
Eq.~(\ref{one}) by roughly $-26\%$, $-18\%$, $-15\%$, and $-16\%$ for
neutrinos, charged leptons, up-type quarks, and down-type quarks,
respectively, i.e., we observe an appreciable screening in all cases.
In comparison, we remark that the relative QCD correction in Eq.~(\ref{drho})
is merely $-10\%$ for the same $\alpha_s$ value.

The examples considered here give support to the heuristic rule that,
in the electroweak on-shell scheme formulated with $G_F$,
the ${\cal O}(G_Fm_t^2)$ terms are screened by their QCD corrections.
By the same token, this screening is weakened---or possibly converted into
antiscreening---when the top-quark mass is renormalized according to the
$\overline{\mbox{MS}}$ scheme, with the choice $\mu={\cal O}(m_t)$.
This may be understood by observing that, for $\mu=m_t$, the
$\overline{\mbox{MS}}$ mass \cite{tar},
\begin{equation}
\label{msmass}
\overline m_t(\mu)=m_t\left[1+C_F{\alpha_s\over\pi}\left({3\over4}
\ln{m_t^2\over\mu^2}-1\right)\right],
\end{equation}
is smaller than $m_t$, the reduction of $x_t$ being approximately $9\%$ for
$\alpha_s=0.108$.
Equation~(\ref{msmass}) follows on from Eq.~(\ref{top}) by discarding the
poles in $\epsilon$.
Consequently, the negative QCD corrections are partly absorbed into $x_t$,
as we pass from the on-shell scheme to the $\overline{\mbox{MS}}$ scheme.
In fact, the relative QCD corrections are then only
$-16\%$ for $\Gamma\left(Z\to\nu\bar\nu H\right)$,
$-9\%$ for $\Gamma\left(Z\to\ell^+\ell^-H\right)$ and $\sigma(e^+e^-\to ZH)$,
$-7\%$ for $\Gamma\left(Z\to d\bar dH\right)$ and $\Gamma(H\to ZZ)$, and
$-6\%$ for $\Gamma\left(Z\to u\bar uH\right)$,
where $\nu$, $\ell$, $u$, and $d$ are generic neutrinos, charged leptons,
up-type quarks, and down-type quarks, respectively.

\section{Conclusions}

The quantum corrections to the production and decay processes of the SM
Higgs boson are now well established in the one-loop approximation \cite{bak}.
Since experiments seem to favour a high-mass top quark \cite{cdf},
one is led to focus attention on the leading high-$m_t$ terms, which are
of ${\cal O}(G_Fm_t^2)$, and one would like to gain control over the dominant
shifts in these terms due to higher-order effects, which are of
${\cal O}(G_F^2m_t^4)$ and ${\cal O}(\alpha_sG_Fm_t^2)$.
Some work in that direction has already been done.
The ${\cal O}(\alpha_sG_Fm_t^2)$ corrections have been evaluated for the
Higgs-boson decays into fermions \cite{hff} and, in particular, into bottom
quarks \cite{hbb,kwi}.
Furthermore, the leading top-quark-induced ${\cal O}(\alpha_s^2)$ corrections
to the Higgs-boson decays into quarks have been found \cite{hqq}.
In this context, we should also mention the ${\cal O}(\alpha_s)$
\cite{daw,hgg} and ${\cal O}(G_Fm_t^2)$ \cite{gam} corrections to the
Higgs-boson decay into gluons, which is mediated by a top-quark loop.

In this article, we continued this research program by presenting the two-loop
\linebreak
${\cal O}(\alpha_sG_Fm_t^2)$ corrections to $\Gamma(H\to ZZ)$,
$\Gamma\left(Z\to f\bar fH\right)$, and $\sigma(e^+e^-\to ZH)$.
As in a previous work \cite{hbb}, we took advantage of a low-energy
theorem, which allowed us to reduce the task of solving two-loop three-point
integrals to a two-loop two-point problem.
Our results are in line with all previous studies of
${\cal O}(\alpha_sG_Fm_t^2)$ corrections to electroweak processes, which have
always shown that such corrections screen the leading $m_t$ dependence of the
one-loop results.
For $\alpha_s=0.108$, the screening effects amount to roughly
$-26\%$ for $\Gamma\left(Z\to\nu\bar\nu H\right)$,
$-18\%$ for $\Gamma\left(Z\to\ell^+\ell^-H\right)$ and $\sigma(e^+e^-\to ZH)$,
$-16\%$ for $\Gamma\left(Z\to d\bar dH\right)$ and $\Gamma(H\to ZZ)$, and
$-15\%$ for $\Gamma\left(Z\to u\bar uH\right)$,
where $\nu$, $\ell$, $u$, and $d$ are generic neutrinos, charged leptons,
up-type quarks, and down-type quarks, respectively.

\end{document}